\documentclass[10pt]{article}

\usepackage{authblk}
\usepackage{natbib}
\usepackage{mathtools}
\usepackage{amsmath}
\usepackage{amsthm}
\usepackage{amssymb}
\usepackage{amsbsy}
\usepackage{amsfonts}
\usepackage{amscd}
\usepackage{mathrsfs}
\usepackage{bm}
\usepackage{breqn}
\usepackage{color, soul}
\usepackage{enumerate}
\usepackage{empheq}
\usepackage{rotating}
\usepackage{ftnxtra}
\usepackage{fnpos}
\usepackage[titletoc,toc,title]{appendix}
\usepackage{euscript}
\usepackage{graphicx}
\usepackage{epsfig}
\usepackage{epstopdf}
\DeclareGraphicsExtensions{.pdf,.png,.jpg,.eps}
\usepackage{pstool}
\usepackage{upgreek}
\usepackage{mathrsfs}

\usepackage{psfrag}

\numberwithin{equation}{section}

\theoremstyle{plain}	
\newtheorem{thm}{Theorem}[section]

\newtheorem*{prop*}{Proposition} 
\theoremstyle{definition}	

\newtheorem{remark}[thm]{Remark}

\usepackage{graphicx}
\usepackage{lipsum}

\usepackage{caption}
\usepackage{subcaption}

\usepackage{hyperref}
\hypersetup{colorlinks=true, linkcolor=blue}
\hypersetup{colorlinks=true,citecolor=blue}

\DeclareMathAlphabet{\mathpzc}{OT1}{pzc}{m}{it}

\usepackage{amsmath, amsthm, amssymb}

\usepackage{cleveref}

\DeclarePairedDelimiter\abs{\lvert}{\rvert}

\makeatletter
\newsavebox{\@brx}
\newcommand{\llangle}[1][]{\savebox{\@brx}{\(\m@th{#1\langle}\)}%
  \mathopen{\copy\@brx\mkern2mu\kern-0.9\wd\@brx\usebox{\@brx}}}
\newcommand{\rrangle}[1][]{\savebox{\@brx}{\(\m@th{#1\rangle}\)}%
  \mathclose{\copy\@brx\mkern2mu\kern-0.9\wd\@brx\usebox{\@brx}}}%
\let\oldabs\abs
\def\abs{\@ifstar{\oldabs}{\oldabs*}}
\makeatother

\usepackage{accents}

    {\end{bmatrix}}%

\usepackage{enumitem}

\usepackage[utf8]{inputenc}
\usepackage{dutchcal}
\usepackage{multirow}

\usepackage{xcolor}

\usepackage{setspace}

\newcommand{\beq}{\begin{equation}}
\newcommand{\eeq}{\end{equation}}
\newcommand{\beqs}{\begin{eqnarray}}
\newcommand{\eeqs}{\end{eqnarray}}
\newcommand{\beql}{\begin{equation} \label}
\newcommand{\half}{\frac{1}{2}}

\newcommand{\p}{\partial}

\newcommand{\scl}{\mathcal{L}}

\newcommand{\s}{\mathsf{s}}

\newcommand{\R}{\mathbb{R}}

\begin{document}

\title{\textbf{Dual Variational Principles for Curl Forces}}

\author[1]{Arash Yavari\thanks{e-mail: arash.yavari@ce.gatech.edu}}
\author[2]{Amit Acharya\thanks{e-mail: acharyaamit@cmu.edu}}
\affil[1]{\small \textit{School of Civil and Environmental Engineering and The George W. Woodruff School of Mechanical Engineering, Georgia Institute of Technology, Atlanta, GA 30332, USA,}}
\affil[2]{\small \textit{Department of Civil \& Environmental Engineering, and Center for Nonlinear Analysis, Carnegie Mellon University, Pittsburgh, PA 15213, USA}}

\maketitle

\begin{abstract}
\noindent Curl forces are position-dependent, non-conservative, and non-dissipative forces that, in general, cannot be derived from an ordinary potential energy. Consequently, their equations of motion do not, in general, follow from a standard variational principle. In this paper, we present a dual variational formulation for particle dynamics under curl forces. By introducing variables dual to position and velocity and an auxiliary function, we construct a pre-dual action in which the equations of motion act as constraints. Stationarity with respect to the primal variables defines a dual-to-primal mapping, whose substitution into the pre-dual action gives an action expressed entirely in terms of the dual variables. The Euler--Lagrange equations of the dual action recover both the original equations of motion and their prescribed initial conditions. We also introduce an auxiliary dual Hamiltonian that is conserved along stationary dual trajectories, although it does not represent the physical energy. The formulation is illustrated using two nonlinear curl force fields in two and three dimensions and the classical Ziegler column. These examples demonstrate that non-conservative curl-force dynamics can admit variational descriptions even in the absence of an ordinary potential energy or a conventional Lagrangian.
\end{abstract}

\begin{description}
\item[Keywords:] Nonconservative force, curl force, dual variational principle, dual-to-primal mapping, auxiliary Hamiltonian, Ziegler column.
\end{description}

\tableofcontents

\section{Introduction}

In classical mechanics, a conservative force field is derived from a potential energy. Most familiar non-conservative forces, including friction and viscous drag, are dissipative. Non-conservativity, however, does not necessarily imply dissipation. An important class of non-conservative, non-dissipative forces consists of position-dependent force fields
\begin{equation}
    \mathbf{F}=\mathbf{F}(\mathbf{x})\,,\qquad    \mathbf{x}\in\mathbb{R}^{n}\,,
    \qquad    n=2,3\,,
\end{equation}
that are independent of velocity such that $\operatorname{curl}\mathbf{F}\neq\mathbf{0}$. Following \citet{Berry2012}, we refer to such force fields as \emph{curl forces}. Position-dependent non-conservative forces are also known as \emph{positional forces}, \emph{pseudo-gyroscopic forces}, or \emph{circulatory forces} \citep{Ziegler1977,Kirillov2021}.

Curl forces are non-conservative but non-dissipative; in particular, their dynamics preserves phase-space volume. Nevertheless, their behavior differs fundamentally from that of conservative systems, and the standard form of Noether's theorem is not directly applicable \citep{Berry2012,Berry2013}. Recent numerical studies reveal behavior ranging from integrable Hamiltonian dynamics to apparently non-Hamiltonian motion with no invariant in phase space \citep{Berry2025Illustrations}. Applications of curl forces in optics and ion trapping were discussed by \citet{Guha2020}. Prominent finite-dimensional mechanical examples arise from follower loads, whose direction changes with the deformation so as to maintain a prescribed orientation relative to the current configuration. Tangential loads applied to flexible beams and columns are classical examples. Such loads have played an important role in stability theory since the 1950s \citep{Pfluger1950,Pfluger1955,Beck1952,Ziegler1952,Ziegler1953,Ziegler1977,Bolotin1963}, with applications in structural mechanics, aeroelasticity, fluid--structure interaction, and rotordynamics. Although the physical realizability of ideal follower loads was long debated \citep{Elishakoff2005,Koiter1996}, experimental realizations have since been reported \citep{Bigoni2011,Bigoni2018,Cazzolli2020}. Curl forces also arise in continuum mechanics, including Cauchy elasticity; their occurrence in continuum theories and the associated literature were recently reviewed by \citet{YavariGoriely2025Cauchy}.

For a conservative force field $\mathbf{F}=\mathbf{F}(\mathbf{x})$, there exists a scalar potential energy $U=U(\mathbf{x})$ such that
\begin{equation}
    \mathbf{F}(\mathbf{x}) = -\nabla U(\mathbf{x}) = -\frac{\partial U}{\partial\mathbf{x}}    \,.
\end{equation}
No such scalar potential energy exists for a curl force. Nevertheless, the absence of an ordinary potential does not imply the absence of a potential-like representation. In a recent paper, \citet{YavariGoriely2025Curl} studied the geometric structure of curl forces using the work $1$-form associated with the force field. 
By applying the Darboux classification of differential $1$-forms, they obtained the following local representations of curl forces:
\begin{subequations}\label{Curl-Force-Representations}
\begin{alignat}{2}
	\mathbf{F}(\mathbf{x})
	&=-V(\mathbf{x})\,\nabla U(\mathbf{x})\,,
	&\qquad& n=2\,,
	\label{Curl-Force-2D}\\
	\mathbf{F}(\mathbf{x})
	&=-V(\mathbf{x})\,\nabla U(\mathbf{x})
	-\nabla W(\mathbf{x})\,,
	&& n=3\,.
	\label{Curl-Force-3D}
\end{alignat}
\end{subequations}
Thus, a two-dimensional curl force requires at most the two generalized potentials $U$ and $V$, whereas a three-dimensional curl force requires at most the three generalized potentials $U$, $V$, and $W$. 
For both representations,
$\operatorname{curl}\mathbf{F}
=-\nabla V\times\nabla U$, where in two dimensions the cross product is understood as its scalar out-of-plane component. Hence, the force is a curl force precisely where
$\nabla V\times\nabla U\neq\mathbf{0}$. 
They also derived equations for calculating these potentials. Determining the generalized potentials requires solving first-order partial differential equations. Although these equations admit local solutions under suitable regularity assumptions, globally defined continuous solutions are not guaranteed in general, and 
the generalized potential $V$ may vanish at some points in its domain.
The same article examined the work performed by curl forces in closed and cyclic motions. In particular, unlike dissipative forces, a curl force performs zero net work over a cyclic motion followed by its reverse. It also investigated the accessibility properties of curl forces and introduced, for every curl force, an associated conservative auxiliary force. The potential representation of curl forces was subsequently generalized to arbitrary space dimensions by \citet{Kycia2025Classification}.

The generalized potentials provide a representation of the original curl force, but do not by themselves yield a Lagrangian for its dynamics. Moreover, the decomposition $\mathbf{F}=\mathbf{F}_{\mathrm{c}}+\mathbf{F}_{\mathrm{nc}}$ into conservative and non-conservative parts is not unique. On any region of configuration space where the generalized potentials are sufficiently regular and $V\neq0$, a chosen decomposition determines an auxiliary conservative force and hence an ordinary local Lagrangian and Hamiltonian for the corresponding auxiliary dynamics, but not for the original curl-force dynamics. When pulled back to the original motion through the nonlocal time-integral relations, the auxiliary Hamiltonian is constant along any trajectory segment on which the construction remains well defined. This (segment-wise) conserved quantity is not the physical energy and depends on the history of the original trajectory rather than only on its instantaneous phase-space state. It therefore does not partition the original phase space into invariant level sets in the usual manner. In contrast, for an arbitrary curl force, the present formulation defines a local Hamiltonian in terms of the dual fields, wherever the corresponding Legendre map is locally invertible, with a pointwise algebraic dual-to-primal mapping.

The absence of an ordinary potential energy raises the question of whether curl-force dynamics can nevertheless be described by a variational principle. 
\citet{Berry2015} showed that a special class of curl forces admits Hamiltonians consisting of an anisotropic kinetic energy and a scalar potential; that construction, however, does not apply to an arbitrary curl force. The underlying theory of quasi-Lagrangian Newton and cofactor systems for nonconservative Newton equations with quadratic first integrals was developed by \citet{RauchWojciechowskiEtAl1999} and \citet{Lundmark2003}. Connections between this framework and planar linear Hamiltonian curl-force systems were studied by \citet{Guha2018}, while \citet{GhoseChoudhuryGuha2019} extended the approach to certain planar quadratic and cubic curl forces.
In this paper, we develop dual variational principles for arbitrary curl forces by introducing variables dual to the position and velocity, together with an auxiliary function, and constructing a pre-dual action involving both the primal and dual variables.  Part of the design of the variational principle involves a dual-to-primal (DtP) mapping, whose substitution into the pre-dual action yields an action expressed entirely in terms of the dual variables. The Euler--Lagrange (E-L) equations of the resulting dual action recover the original equations of motion and the prescribed initial conditions. The formulation is illustrated using two nonlinear curl forces and the classical Ziegler column. Our dual variational principles are based on constructing dual Lagrangians. We also construct auxiliary dual Hamiltonians, albeit not single-valued in general, for the developed Lagrangians; these Hamiltonians are not necessary for the existence of the variational principles.

The technique employed to generate the dual variational principles follows the general considerations related to Newtonian particle mechanics, including dissipative forces and anholonomic constraints, developed in \citep{Ach6}, with application to a particle chain in \citep{ach_cmds}. Applications of the approach to various nonlinear PDE theories are demonstrated in  \citep{acharyaQAM,ach2,ASZ24,AGS24,Achmhd,sga}, with the last work including dual variational principles for finite Cauchy elasticity. Dual variational minimum principles for  possibly heterogeneous, linear, Cauchy elastodynamics are developed in \citep{acharya2026new}. This is a surprising result since it is well-understood that elastodynamics in its primal variables, and in the presence of a strain energy function, at most provides an extremal principle. While still in its infancy, encouraging progress has also been demonstrated in approximating solutions of nonlinear differential equations related to continuum mechanics in  \citep{ka1,sga,ka2,kpa}. These cases addressed Euler's nonlinear system of ODEs for the dynamics of a rigid body, nonconvex elastodynamics and statics of a bar (without higher gradient regularization), the (inviscid) Burgers equation, and the problem of traveling waves of a dispersive, nonlocal, nonlinear semi-discrete Burgers equation. Furthermore, \cite{AG_control} show the application of the method to problems of Optimal Control, and \cite{AV_nash} to the Nash system of Deterministic Game Theory.

This paper is organized as follows. In \S\ref{Sec:DualVariationalFormulation}, we introduce the dual variables, construct the dual action, and derive the corresponding auxiliary dual Hamiltonian for arbitrary curl forces. In \S\ref{Sec:Examples}, we apply the dual variational formulation to two nonlinear curl force fields and the classical Ziegler column. Some concluding remarks are given in \S\ref{Sec:Conclusions}.

\section{A dual variational formulation for curl forces}\label{Sec:DualVariationalFormulation}

Let $\mathcal{Q}$ be the finite-dimensional configuration space of the particle ($\mathcal{Q}=\mathbb{R}^n$, $n=2$ or $3$). Its state is denoted by $\boldsymbol{z}=(\boldsymbol{x},\boldsymbol{v})\in T\mathcal{Q}$, where $\boldsymbol{x}\in\mathcal{Q}$ is the position of the particle and $\boldsymbol{v}\in T_{\boldsymbol{x}}\mathcal{Q}$ is its velocity. Thus, $T\mathcal{Q}$ is the state space of the particle. The corresponding dual state is denoted by $\boldsymbol{\zeta}=(\boldsymbol{\xi},\boldsymbol{\eta})\in T_{\boldsymbol{z}}^*(T\mathcal{Q})$. Consequently, $(\boldsymbol{z},\boldsymbol{\zeta})\in T^*(T\mathcal{Q})$. In local coordinates $(x^i,v^i)$ on $T\mathcal{Q}$, the dual state has the representation $\boldsymbol{\zeta}=\xi_i\,\mathrm{d}x^i	+ \eta_i\,\mathrm{d}v^i$. Hence, $\boldsymbol{\xi}$ is dual to the position coordinates $\boldsymbol{x}$, whereas $\boldsymbol{\eta}$ is dual to the velocity coordinates $\boldsymbol{v}$. The variables $\boldsymbol{x}$ and $\boldsymbol{v}$ are tangent-state variables, while $\boldsymbol{\xi}$ and $\boldsymbol{\eta}$ are cotangent-state variables. We refer to $\boldsymbol{z}$ as the \textit{state} and to $\boldsymbol{\zeta}$ as the \textit{dual state}. The natural pairing between the state $\boldsymbol{z}$ and the dual state $\boldsymbol{\zeta}$ is given by
\begin{equation}
	\left\langle \boldsymbol{\zeta},\boldsymbol{z}\right\rangle=
	\left\langle (\boldsymbol{\xi},\boldsymbol{\eta}), (\boldsymbol{x},\boldsymbol{v})\right\rangle
	=\boldsymbol{\xi}\cdot\boldsymbol{x}+\boldsymbol{\eta}\cdot\boldsymbol{v}=\xi_i x^i+\eta_i v^i\,.
\end{equation}

Let us assume that the particle is subject to the curl force $\mathbf{F}(\mathbf{x})$, with initial conditions
$\mathbf{x}(0)=\mathbf{x}_0$ and $\mathbf{v}(0)=\mathbf{v}_0$. Its motion is governed by the following initial-value problem (IVP):
\begin{equation} \label{eq:primal}
\begin{dcases}
	\dot{\mathbf{x}}-\mathbf{v}=\mathbf{0}\,,\\
	\dfrac{d}{dt}(m\mathbf{v})-\mathbf{F}(\mathbf{x})=\mathbf{0}\,,\\
	\mathbf{x}(0)=\mathbf{x}_0\,,\quad
	\mathbf{v}(0)=\mathbf{v}_0\,.
\end{dcases}
\end{equation}

\begin{remark}
The formulation extends directly to a finite system of particles. In this case, $\mathbf{x}$ and $\mathbf{v}$ collect the positions and velocities of all the particles, $m$ is replaced by the corresponding mass matrix $\mathbf{M}$, and $\mathbf{F}(\mathbf{x})$ may include both external and interaction forces. The subsequent dual construction remains unchanged, with $m\mathbf{v}$ replaced by $\mathbf{M}\mathbf{v}$.
\end{remark}

\subsection{A dual action principle}\label{sec:dual_action}

Pairing $\boldsymbol{\xi}$ with the definition of velocity and $\boldsymbol{\eta}$ with the balance of linear momentum motivates the
introduction of the following pre-dual Lagrangian
\begin{equation} \label{PreDualLagrangian}
	\mathcal{L} \big(\mathbf{x},\mathbf{v},\boldsymbol{\xi},\boldsymbol{\eta},
	\dot{\boldsymbol{\xi}},\dot{\boldsymbol{\eta}};\bar{\mathbf{x}},\bar{\mathbf{v}} \big)
	= -\mathbf{x}\cdot\dot{\boldsymbol{\xi}}-m\mathbf{v}\cdot\dot{\boldsymbol{\eta}}
	-\boldsymbol{\xi}\cdot\mathbf{v}	-\boldsymbol{\eta}\cdot\mathbf{F}(\mathbf{x})
	-H\big(\mathbf{x}-\bar{\mathbf{x}},\mathbf{v}-\bar{\mathbf{v}} \big)\,,
\end{equation}
where $H$ is, for the moment, an unspecified smooth function, and $\bar{\mathbf{x}}=\bar{\mathbf{x}}(t)$ and $\bar{\mathbf{v}}=\bar{\mathbf{v}}(t)$ are prescribed position and velocity fields, referred to as \emph{base states}.  The pre-dual action is defined as
\begin{equation}\label{PreDualAction}
	\widehat{S}[\mathbf{x},\mathbf{v},\boldsymbol{\xi},\boldsymbol{\eta}]
	=\int_0^T \mathcal{L} \big(	\mathbf{x},\mathbf{v},\boldsymbol{\xi},\boldsymbol{\eta},
	\dot{\boldsymbol{\xi}},\dot{\boldsymbol{\eta}};\bar{\mathbf{x}},\bar{\mathbf{v}} \big)\,\mathrm{d}t
	-\mathbf{x}_0\cdot\boldsymbol{\xi}(0)	-m\mathbf{v}_0\cdot\boldsymbol{\eta}(0)\,.
\end{equation}
The terminal values $\boldsymbol{\xi}(T)$ and $\boldsymbol{\eta}(T)$ are prescribed; they may, in particular, be chosen to vanish.

We now eliminate the primal variables $\mathbf{x}$ and $\mathbf{v}$ to derive a dual functional with the desired properties. For this, we require that the set of equations $\frac{\p \scl}{\partial\mathbf{x}} = \mathbf{0}$ and $\frac{\p \scl}{\p \mathbf{v}} = \mathbf{0}$, written explicitly as
\begin{equation}\label{eq:DtPEquations}
\begin{dcases}
	\rule{0pt}{5ex}
	\frac{\partial H}{\partial\mathbf{x}}
	\big(\mathbf{x}-\bar{\mathbf{x}},\mathbf{v}-\bar{\mathbf{v}}\big)
	+ \dot{\boldsymbol{\xi}}
	+ \big[D\mathbf{F}(\mathbf{x})\big]^{\mathsf T}\cdot\boldsymbol{\eta} = \mathbf{0}\,,
	\\[1ex]
	\rule[-2.5ex]{0pt}{5ex}
	\frac{\partial H}{\partial\mathbf{v}}
	\big(\mathbf{x}-\bar{\mathbf{x}},\mathbf{v}-\bar{\mathbf{v}}\big)
	+m\dot{\boldsymbol{\eta}}+\boldsymbol{\xi} = \mathbf{0}\,,
\end{dcases}
\end{equation}
be solved for $\mathbf{x}, \mathbf{v}$ in terms of the remaining arguments of $\mathcal{L}$. We assume that $H$ is chosen such that \eqref{eq:DtPEquations} can be solved, at least locally, for $\mathbf{x}$ and $\mathbf{v}$. This defines the dual-to-primal (DtP) mapping
\begin{equation}\label{DtPMap}
	(\mathbf{x},\mathbf{v})=\big(\widehat{\mathbf{x}},\widehat{\mathbf{v}}\big)
	\big(\boldsymbol{\xi},\boldsymbol{\eta},\dot{\boldsymbol{\xi}},\dot{\boldsymbol{\eta}}
	;\bar{\mathbf{x}},\bar{\mathbf{v}} \big)\,.
\end{equation}
Substitution of the dual-to-primal mapping into \eqref{PreDualAction} defines the dual action
\begin{equation} \label{DualAction}
	S[\boldsymbol{\xi},\boldsymbol{\eta}]	
	=\int_0^T \mathcal{L} \big(	\widehat{\mathbf{x}},	\widehat{\mathbf{v}},	\boldsymbol{\xi},\boldsymbol{\eta},
	\dot{\boldsymbol{\xi}},\dot{\boldsymbol{\eta}};\bar{\mathbf{x}},\bar{\mathbf{v}} \big)\,\mathrm{d}t
	-\mathbf{x}_0\cdot\boldsymbol{\xi}(0)	-m\mathbf{v}_0\cdot\boldsymbol{\eta}(0)\,.
\end{equation}
Since the dual-to-primal mapping satisfies $\partial\mathcal{L}/\partial\mathbf{x}=\mathbf{0}$ and $\partial\mathcal{L}/\partial\mathbf{v}=\mathbf{0}$, the dependence of $\widehat{\mathbf{x}}$ and $\widehat{\mathbf{v}}$ on the dual variables does not contribute to the first variation of \eqref{DualAction}. Consequently,
\begin{equation}
	\delta S= \int_0^T
	\left[	-\widehat{\mathbf{x}}\cdot	\delta\dot{\boldsymbol{\xi}}-\widehat{\mathbf{v}}\cdot
	\delta\boldsymbol{\xi} -m\widehat{\mathbf{v}}\cdot \delta\dot{\boldsymbol{\eta}}	
	-\mathbf{F}(\widehat{\mathbf{x}})\cdot\delta\boldsymbol{\eta} \right]\mathrm{d}t
	-\mathbf{x}_0\cdot\delta\boldsymbol{\xi}(0)-m\mathbf{v}_0\cdot\delta\boldsymbol{\eta}(0) \,.
\end{equation}
Integrating by parts and using $\delta\boldsymbol{\xi}(T)=\delta\boldsymbol{\eta}(T)=\mathbf{0}$ gives us
\begin{equation}
	\delta S=	\int_0^T \left[\big(\dot{\widehat{\mathbf{x}}}-\widehat{\mathbf{v}}\big)
	\cdot\delta\boldsymbol{\xi}
	+\big(m\dot{\widehat{\mathbf{v}}} -\mathbf{F}(\widehat{\mathbf{x}})\big)
	\cdot\delta\boldsymbol{\eta}\right]\mathrm{d}t
	+\left[\widehat{\mathbf{x}}(0)-\mathbf{x}_0	\right]\cdot\delta\boldsymbol{\xi}(0)
	+m\left[\widehat{\mathbf{v}}(0)-\mathbf{v}_0 \right]\cdot\delta\boldsymbol{\eta}(0) \,.
\end{equation}
Therefore, the Euler--Lagrange equations and natural initial conditions associated with the dual action are
\begin{equation}\label{eq:dual_EL}
\begin{dcases}
	\dot{\widehat{\mathbf{x}}} -\widehat{\mathbf{v}}=\mathbf{0}\,,	\\
	m\dot{\widehat{\mathbf{v}}}-\mathbf{F}(\widehat{\mathbf{x}})=\mathbf{0}\,,	\\
	\widehat{\mathbf{x}}(0)=\mathbf{x}_0\,, \quad	\widehat{\mathbf{v}}(0)=\mathbf{v}_0\,.
\end{dcases}
\end{equation}
Thus, every stationary point of the dual action, \emph{for any choice of the auxiliary potential $H$} that allows the definition of a DtP map, generates, through the dual-to-primal mapping, a solution of the original initial-value problem.

A sufficient local condition for the existence of the DtP map is the nonsingularity of the Jacobian of the left-hand side of \eqref{eq:DtPEquations} (or the Hessian of $\mathcal{L}$) with respect to
$(\mathbf{x},\mathbf{v})$, and the satisfaction of the equations at one point of the local neighborhood. Strict convexity of $H$ facilitates this invertibility, although it is not by itself sufficient for arbitrary
values of $\boldsymbol{\eta}$ when $\mathbf{F}$ is nonlinear and if the auxiliary potential is restricted to one fixed choice. As explained in detail in \citep{ach3,AV_nash}, due to the Lagrangian $\scl$ being necessarily affine in the dual variables, auxiliary potentials $H$ can always be constructed so that the DtP map equations \eqref{eq:DtPEquations} are satisfied at $(\mathbf{x},\mathbf{v})=(\bar{\mathbf{x}},\bar{\mathbf{v}})$ when $(\boldsymbol{\eta},\dot{\boldsymbol{\eta}}, \boldsymbol{\xi},\dot{\boldsymbol{\xi}}) = (\mathbf{0},\mathbf{0},\mathbf{0},\mathbf{0})$. 
This feature endows an important global-in-time consistency property to the dual formulation---it shows that for any solution to the IVP (for its whole class of initial conditions), there exists at least one dual functional which admits an extremal whose DtP mapped image is given by the solution to the IVP being considered. That dual functional is obtained by the type of $H$ just discussed, with the base state chosen to be the primal solution, and the corresponding dual extremal given by $t \mapsto (\boldsymbol{\xi}(t) , \boldsymbol{\eta}(t) ) = (\mathbf{0},\mathbf{0})$. An important practical corollary of this argument is that for a base state close to a primal solution, the preceding $\mathbf{0}$-dual state serves as a good initial guess for obtaining a dual extremal.  Furthermore, the liberty in choosing the base states and the auxiliary function $H$ can be well-exercised to not only have a well-defined DtP map whenever necessary in a solution procedure for obtaining a solution to the nonlinear equations \eqref{eq:dual_EL}, but even do so utilizing a sequence of \emph{convex} dual variational principles parametrized by base states, see, e.g., \cite[Sec.~3]{AV_nash}.

Finally, it is worth noting here that Hamilton's celebrated action principle does not recover the full set of initial conditions of an initial-value problem (IVP), and in addition requires an acausal boundary condition at final time $T$ to be prescribed, typically on the position (cf., \cite{gurt}). This is an issue of some recent interest \citep{gall,rothkopf2026variational}, a remedy involving forward and backward in time paths for a given initial-value problem. In the present dual formulation, no backward in time evolution is required. Instead, final-time, Dirichlet boundary conditions are prescribed on the dual variables, and due to the necessary presence of the time-derivatives of the dual variables in the DtP map by design of the Lagrangian for a first-order initial-value problem, such a final-time specification does not obstruct the recovery of the causal result for the primal variables at time $T$. For instance, the values of $\dot{\boldsymbol{\xi}}(T)$ and $ \dot{\boldsymbol{\eta}}(T)$ in the dual solution adjust for the specification of $\boldsymbol{\xi}(T)$ and $ \boldsymbol{\eta}(T)$ to recover the correct solution for $\widehat{\mathbf{x}}(T)$ and $\widehat{\mathbf{v}}(T)$, when there is uniqueness of solutions to the primal IVP \eqref{eq:primal}(see solved examples in, e.g., \cite{ka1,suku_ach}). Even in the absence of uniqueness in the primal IVP, any solution to the dual E-L system always generates a genuine solution to the primal problem through the corresponding DtP map.

\subsection{An auxiliary dual Hamiltonian}

Let $\mathcal{L}_{\mathrm{d}}\big(\boldsymbol{\xi},\boldsymbol{\eta},\dot{\boldsymbol{\xi}},\dot{\boldsymbol{\eta}};t\big)=\mathcal{L}\big(\widehat{\mathbf{x}},\widehat{\mathbf{v}},
\boldsymbol{\xi},\boldsymbol{\eta},\dot{\boldsymbol{\xi}},\dot{\boldsymbol{\eta}};\bar{\mathbf{x}},\bar{\mathbf{v}}\big)$ denote the reduced dual Lagrangian. Since the dual-to-primal mapping satisfies
$\left.\frac{\partial\mathcal{L}}{\partial\mathbf{x}}\right|_{(\mathbf{x},\mathbf{v})=(\widehat{\mathbf{x}},\widehat{\mathbf{v}})}=\mathbf{0}$ and $\left.\frac{\partial\mathcal{L}}{\partial\mathbf{v}}\right|_
{(\mathbf{x},\mathbf{v})=(\widehat{\mathbf{x}},\widehat{\mathbf{v}})}=\mathbf{0}$, its canonical momenta are
\begin{subequations}
\begin{align}
	\mathbf{p}_{\boldsymbol{\xi}}
	&=\frac{\partial\mathcal{L}_{\mathrm{d}}}{\partial\dot{\boldsymbol{\xi}}}
	=-\widehat{\mathbf{x}}\,,\\
	\mathbf{p}_{\boldsymbol{\eta}}
	&=\frac{\partial\mathcal{L}_{\mathrm{d}}}{\partial\dot{\boldsymbol{\eta}}}
	=-m\widehat{\mathbf{v}}\,.
\end{align}
\end{subequations}
The Legendre map associated with $\mathcal{L}_{\mathrm{d}}$ is $\big(\boldsymbol{\xi},\boldsymbol{\eta},\dot{\boldsymbol{\xi}},\dot{\boldsymbol{\eta}};t\big)
\mapsto\big(\boldsymbol{\xi},\boldsymbol{\eta},\mathbf{p}_{\boldsymbol{\xi}},\mathbf{p}_{\boldsymbol{\eta}};t\big)$. Whenever this map is locally invertible at a specified point
$(\dot{\boldsymbol{\xi}},\dot{\boldsymbol{\eta}})$ in the dual-velocity space, the inverse function theorem guarantees the existence of a unique local inverse in a neighborhood of the corresponding point
$(\mathbf{p}_{\boldsymbol{\xi}},\mathbf{p}_{\boldsymbol{\eta}})$ in the dual-momentum space.
Thus, after selecting the corresponding local branch, the dual velocities can be expressed as follows:
\begin{equation}
	\dot{\boldsymbol{\xi}}
	 = \mathbf{V}_{\boldsymbol{\xi}}
	 \big(\boldsymbol{\xi},\boldsymbol{\eta},
	 \mathbf{p}_{\boldsymbol{\xi}},
	 \mathbf{p}_{\boldsymbol{\eta}};t\big),
	\qquad
	\dot{\boldsymbol{\eta}}
	 = \mathbf{V}_{\boldsymbol{\eta}}
	 \big(\boldsymbol{\xi},\boldsymbol{\eta},
	 \mathbf{p}_{\boldsymbol{\xi}},
	 \mathbf{p}_{\boldsymbol{\eta}};t\big)\,.
\end{equation}
If several velocity points yield the same momenta, each such point may determine a different local inverse and hence a different local branch of the Hamiltonian. Local invertibility alone therefore does not define a globally single-valued Hamiltonian. 

Positive definiteness everywhere of the Hessian of $\mathcal{L}_{\mathrm{d}}$ with respect to
$(\dot{\boldsymbol{\xi}},\dot{\boldsymbol{\eta}})$ implies strict convexity in the velocities and hence global injectivity of the Legendre map. An additional condition such as superlinear growth in the velocities
gives surjectivity onto the entire momentum space. When convexity fails, the Legendre--Fenchel transform may eliminate dynamically important nonconvex features of the primal problem inherited by the dual Lagrangian, such as energy barriers and saddle points, and is therefore not considered here.

The independent variables of a dual Hamiltonian are therefore $\boldsymbol{\xi}$, $\boldsymbol{\eta}$,
$\mathbf{p}_{\boldsymbol{\xi}}$, and $\mathbf{p}_{\boldsymbol{\eta}}$, together with $t$ when the reduced dual Lagrangian depends explicitly on time. A dual Hamiltonian is defined as
\begin{equation}
	\mathscr{H}_{\mathrm{d}}
	\big(\boldsymbol{\xi},\boldsymbol{\eta},
	\mathbf{p}_{\boldsymbol{\xi}},\mathbf{p}_{\boldsymbol{\eta}};t\big)
	=\mathbf{p}_{\boldsymbol{\xi}}\cdot\mathbf{V}_{\boldsymbol{\xi}}
	+\mathbf{p}_{\boldsymbol{\eta}}\cdot\mathbf{V}_{\boldsymbol{\eta}}\\
	-\mathcal{L}_{\mathrm{d}}\big(\boldsymbol{\xi},\boldsymbol{\eta},
	\mathbf{V}_{\boldsymbol{\xi}},\mathbf{V}_{\boldsymbol{\eta}};t\big)\,.
\end{equation}
Using the definitions of the canonical momenta and the pre-dual Lagrangian, the dual Hamiltonian is simplified to read
\begin{equation}\label{eq:DualHamiltonian}
	\mathscr{H}_{\mathrm{d}}
	=\boldsymbol{\xi}\cdot\widehat{\mathbf{v}}
	+\boldsymbol{\eta}\cdot\mathbf{F}(\widehat{\mathbf{x}})
	+H\big(\widehat{\mathbf{x}}-\bar{\mathbf{x}},
	\widehat{\mathbf{v}}-\bar{\mathbf{v}}\big)\,,
\end{equation}
where $\widehat{\mathbf{x}}$ and $\widehat{\mathbf{v}}$ are understood as functions of
$\big(\boldsymbol{\xi},\boldsymbol{\eta},
\mathbf{p}_{\boldsymbol{\xi}},\mathbf{p}_{\boldsymbol{\eta}};t\big)$
through the inverse Legendre map. The corresponding Hamilton equations are
\begin{equation}\label{eq:DualHamiltonEquations}
\begin{dcases}
	\dot{\boldsymbol{\xi}}
	=\dfrac{\partial\mathscr{H}_{\mathrm{d}}}{\partial\mathbf{p}_{\boldsymbol{\xi}}}\,,
	&\dot{\mathbf{p}}_{\boldsymbol{\xi}}
	=-\dfrac{\partial\mathscr{H}_{\mathrm{d}}}{\partial\boldsymbol{\xi}}\,,
	\\[2ex]
	\dot{\boldsymbol{\eta}}
	=\dfrac{\partial\mathscr{H}_{\mathrm{d}}}{\partial\mathbf{p}_{\boldsymbol{\eta}}}\,,
	&\dot{\mathbf{p}}_{\boldsymbol{\eta}}
	=-\dfrac{\partial\mathscr{H}_{\mathrm{d}}}{\partial\boldsymbol{\eta}}\,.
\end{dcases}
\end{equation}
Along a solution of \eqref{eq:DualHamiltonEquations}, for as long as the solution remains within the domain of definition of the selected locally defined dual Hamiltonian, its total time derivative is computed as
\begin{equation}
\begin{aligned}
	\frac{\mathrm{d}}{\mathrm{d}t}\mathscr{H}_{\mathrm{d}}
	={}&
	\frac{\partial\mathscr{H}_{\mathrm{d}}}{\partial\boldsymbol{\xi}}
	\cdot\dot{\boldsymbol{\xi}}
	+\frac{\partial\mathscr{H}_{\mathrm{d}}}{\partial\boldsymbol{\eta}}
	\cdot\dot{\boldsymbol{\eta}}
	+\frac{\partial\mathscr{H}_{\mathrm{d}}}{\partial\mathbf{p}_{\boldsymbol{\xi}}}
	\cdot\dot{\mathbf{p}}_{\boldsymbol{\xi}}
	+\frac{\partial\mathscr{H}_{\mathrm{d}}}{\partial\mathbf{p}_{\boldsymbol{\eta}}}
	\cdot\dot{\mathbf{p}}_{\boldsymbol{\eta}}
	+\frac{\partial\mathscr{H}_{\mathrm{d}}}{\partial t} \\
	={}&
	-\dot{\mathbf{p}}_{\boldsymbol{\xi}}\cdot\dot{\boldsymbol{\xi}}
	-\dot{\mathbf{p}}_{\boldsymbol{\eta}}\cdot\dot{\boldsymbol{\eta}}
	+\dot{\boldsymbol{\xi}}\cdot\dot{\mathbf{p}}_{\boldsymbol{\xi}}
	+\dot{\boldsymbol{\eta}}\cdot\dot{\mathbf{p}}_{\boldsymbol{\eta}}
	+\frac{\partial\mathscr{H}_{\mathrm{d}}}{\partial t}
	=\frac{\partial\mathscr{H}_{\mathrm{d}}}{\partial t}\,.
\end{aligned}
\end{equation}
Consider a time interval on which the dual Hamiltonian (evaluated along the considered solution) is well defined
in a neighborhood, in momentum space, of the momenta corresponding under
the Legendre map to the initial velocity for that interval. If
$\bar{\mathbf{x}}$ and $\bar{\mathbf{v}}$ are independent of time and
$\mathbf{F}$ does not depend explicitly on time, then
$\partial\mathscr{H}_{\mathrm{d}}/\partial t=0$, and hence
\begin{equation}
	\frac{\mathrm{d}}{\mathrm{d}t}\mathscr{H}_{\mathrm{d}}=0\,,
\end{equation}
along dual Hamiltonian trajectories (within the considered neighborhood of momentum space). This conserved quantity is an auxiliary dual Hamiltonian; in general, it is not the physical energy of
the particle.

Since $H$ is freely chosen, on each sufficiently short time interval, the base states defining it may be chosen to be constant in time and taking values of the primal position and velocity at the beginning of that interval. In a small time interval, it is natural to expect to be able to find a dual extremal 
with such an auxiliary potential $H$, given the corollary to the consistency property of the dual formulation discussed in Sec.~\ref{sec:dual_action}. By restarting the initial-value problem, using the flow property, and exploiting the freedom in the choice of $H$, one thereby obtains a large family of concatenated-in-time, locally defined, explicitly time-independent dual Lagrangians and corresponding Hamiltonians, each of which is conserved along the corresponding dual trajectory segment and, through the DtP map, a segment of the corresponding primal solution. 
Thus, even for dissipative or non-conservative primal systems that do not have a conserved physical energy, auxiliary dual Hamiltonians that are conserved piecewise in time are readily available and may provide useful tools for the analysis and computation of different aspects of the primal problem.

\section{Examples}\label{Sec:Examples}

In this section, we illustrate the dual variational formulation developed in the previous section using two nonlinear curl force fields and the classical Ziegler column. These examples demonstrate how the structure of the force field determines the dual-to-primal mapping. In all the following examples, we will employ the zero base state, which results in the DtP mapping not being uniformly valid for all possible dual states $t \mapsto (\boldsymbol{\xi}(t),\boldsymbol{\eta}(t))$. In \citep[Sec.~3]{AV_nash}, and \citep{ka1,ka2,kpa} it is shown how base state `resets' can be utilized in rigorous as well as algorithmic contexts to always operate with a well-defined DtP map.

\subsection{A two-dimensional nonlinear curl force}

Let us apply the dual variational formulation to the following two-dimensional curl force that was considered in \citep{YavariGoriely2025Curl}
\begin{equation}
    \mathbf{F}(x,y) = -\frac{F_0}{a^3}\big(xy^2,x^3\big)\,.
\end{equation}
For convenience, let $k=\frac{F_0}{a^3}$, and choose the base state $\bar{\mathbf{x}}=\bar{\mathbf{v}}=\mathbf{0}$. The Jacobian of the force field is
\begin{equation}
    D\mathbf{F}(x,y)=
    -k  \begin{bmatrix}
        y^2 & 2xy\\
        3x^2 & 0
    \end{bmatrix}\,.
\end{equation}
Let $H=H(\mathbf{x},\mathbf{v})$ be a smooth function, and write the dual variables as $\boldsymbol{\xi}=(\xi_x,\xi_y)$ and $\boldsymbol{\eta}=(\eta_x,\eta_y)$. The dual-to-primal equations \eqref{eq:DtPEquations} take the form
\begin{subequations}\label{BerryGeneralDualToPrimal}
\begin{align}
    \frac{\partial H}{\partial x} &=-\dot{\xi}_x+k\big(y^2\,\eta_x +3x^2\,\eta_y \big)\,, \\
    \frac{\partial H}{\partial y} &=-\dot{\xi}_y+2k\,xy\,\eta_x\,, \\
    \frac{\partial H}{\partial v_x} &=-m\dot{\eta}_x-\xi_x\,, \\
    \frac{\partial H}{\partial v_y} &=-m\dot{\eta}_y-\xi_y\,.
\end{align}
\end{subequations}
Provided that these equations can be solved locally for $(x,y,v_x,v_y)$, they define the dual-to-primal mapping $(\mathbf{x},\mathbf{v})=(\widehat{\mathbf{x}},\widehat{\mathbf{v}})  \big(\boldsymbol{\xi},\boldsymbol{\eta},\dot{\boldsymbol{\xi}},\dot{\boldsymbol{\eta}}\big)$. In order to obtain a more explicit representation, let us consider the quadratic function
\begin{equation}
    H(\mathbf{x},\mathbf{v})= \frac{\alpha}{2}\big(x^2+y^2\big)+ \frac{\beta}{2}\big(v_x^2+v_y^2\big)\,,
    \qquad    \alpha,\beta>0\,.
\end{equation}
The dual-to-primal equations \eqref{BerryGeneralDualToPrimal} are then simplified to read
\begin{subequations}\label{BerryDualToPrimal}
\begin{align}
    \alpha x
    &= -\dot{\xi}_x
    +k\big(y^2\eta_x+3x^2\eta_y\big)\,,
    \label{BerryDualToPrimal-a}\\
    \alpha y
    &= -\dot{\xi}_y+2kxy\eta_x\,,
    \label{BerryDualToPrimal-b}\\
    \beta v_x
    &= -m\dot{\eta}_x-\xi_x\,,
    \label{BerryDualToPrimal-c}\\
    \beta v_y
    &= -m\dot{\eta}_y-\xi_y\,.
    \label{BerryDualToPrimal-d}
\end{align}
\end{subequations}
Thus, the velocity part of the dual-to-primal mapping reads
\begin{equation} \label{v-hat-Ex1}
    \widehat{\mathbf{v}}= -\frac{1}{\beta} \big(m\,\dot{\boldsymbol{\eta}}+\boldsymbol{\xi}\big)\,.
\end{equation}
Eq.~\eqref{BerryDualToPrimal-b} gives us
\begin{equation} \label{y-hat-Ex1}
    \widehat{y} =  -\frac{\dot{\xi}_y}{\alpha-2k\,\eta_x\,\widehat{x}}\,,
\end{equation}
provided that $\alpha-2k\,\eta_x\,\widehat{x}\neq0$. 
Substituting into \eqref{BerryDualToPrimal-a} gives us the following algebraic equation for $\widehat{x}$:
\begin{equation}\label{BerryDualAlgebraicEquation}
    \alpha\,\widehat{x}+\dot{\xi}_x -3k\,\eta_y\,\widehat{x}^{\,2}
    -\frac{k\,\eta_x\,\dot{\xi}_y^{\,2}}{\big(\alpha-2k\,\eta_x\,\widehat{x}\big)^2}
    =0\,.
\end{equation}
Therefore, any locally selected root of \eqref{BerryDualAlgebraicEquation}, together with \eqref{v-hat-Ex1} and \eqref{y-hat-Ex1} defines a local dual-to-primal map. 
Eqs.~\eqref{BerryDualToPrimal-a} and \eqref{BerryDualToPrimal-b} can be written as $\mathbf{G}(x,y)=\mathbf{0}$, where
\begin{equation}
	\mathbf{G}(x,y)=
	\begin{bmatrix}
		\alpha x+\dot{\xi}_x-k\eta_x\,y^2-3k\eta_y\,x^2\\
		\alpha y+\dot{\xi}_y-2k\eta_x\,xy
	\end{bmatrix}\,.
\end{equation}
In calculating the Jacobian with respect to $(x,y)$, the dual variables $\boldsymbol{\xi}$, $\boldsymbol{\eta}$ and their time derivatives are held fixed. Thus, in particular, $\partial\dot{\xi}_i/\partial x=\partial\dot{\xi}_i/\partial y=0$. The Jacobian of $\mathbf{G}$ with respect to $(x,y)$ is written as
\begin{equation}
	D_{(x,y)}\mathbf{G}(x,y)=
	\begin{bmatrix}
		\alpha-6k\eta_y\,x & -2k\eta_x\,y\\
		-2k\eta_x\,y & \alpha-2k\eta_x\,x
	\end{bmatrix}\,.
\end{equation}
Since $\beta>0$, the velocity part of the dual-to-primal mapping is already invertible. Therefore, by the implicit function theorem, the complete dual-to-primal mapping is locally well-defined whenever
\begin{equation}
	\begin{vmatrix}
		\alpha-6k\eta_y\,\widehat{x} & -2k\eta_x\,\widehat{y}\\
		-2k\eta_x\,\widehat{y} & \alpha-2k\eta_x\,\widehat{x}
	\end{vmatrix}\neq0\,.
\end{equation}

The corresponding reduced dual Lagrangian is
\begin{equation}
	\mathcal{L}_{\mathrm d} = -\widehat{\mathbf{x}}\cdot\dot{\boldsymbol{\xi}} 
	-m\,\widehat{\mathbf{v}}\cdot\dot{\boldsymbol{\eta}}
	-\boldsymbol{\xi}\cdot\widehat{\mathbf{v}}+k\,\eta_x\,\widehat{x}\,\widehat{y}^{\,2}
	+k\,\eta_y\,\widehat{x}^{\,3}-\frac{\alpha}{2}  \big(\widehat{x}^{\,2}+\widehat{y}^{\,2}\big) 
	- \frac{\beta}{2}\big(\widehat{v}_x^{\,2}+\widehat{v}_y^{\,2}\big)\,.
\end{equation}
Stationarity of the associated dual action recovers
\begin{equation}
\begin{dcases}
    \dot{\widehat{x}}=\widehat{v}_x\,,  & \dot{\widehat{y}}=\widehat{v}_y\,,   \\[1ex]
    m\dot{\widehat{v}}_x=-k\widehat{x}\widehat{y}^{\,2}\,, & m\dot{\widehat{v}}_y=-k\widehat{x}^{\,3}\,,
\end{dcases}
\end{equation}
together with the prescribed initial conditions. Hence, although this curl force does not possess an ordinary potential energy, its initial-value problem follows from a variational principle expressed entirely in terms of the dual variables. Notice that this construction uses the force field directly.

\begin{remark}[Quadratic lifting]
An explicit dual-to-primal map can alternatively be obtained by introducing the auxiliary primal variables $r=y^2$ and $s=x^2$. In terms of the enlarged set of primal variables, the force field reads
\begin{equation}
	\mathbf{F}(x,r,s)=-k\big(xr,xs\big)\,.
\end{equation}
The original initial-value problem is then equivalent to the following differential-algebraic system
\begin{equation}
\begin{dcases}
	\dot{x}-v_x=0\,, & \dot{y}-v_y=0\,,\\[1ex]
	m\dot{v}_x+kxr=0\,, & m\dot{v}_y+kxs=0\,,\\[1ex]
	y^2-r=0\,, & x^2-s=0\,.
\end{dcases}
\end{equation}
All the nonlinearities in this enlarged system are quadratic. Let $\lambda$ and $\mu$ denote the dual variables associated with the algebraic constraints $y^2-r=0$ and $x^2-s=0$, respectively. The augmented pre-dual Lagrangian is written as\footnote{This Lagrangian is obtained by pairing $\xi_x$ and $\xi_y$ with the kinematic equations, $\eta_x$ and $\eta_y$ with the momentum equations, and $\lambda$ and $\mu$ with the algebraic constraints. Integration by parts transfers the time derivatives of $x$, $y$, $v_x$, and $v_y$ to the corresponding dual variables; the resulting boundary terms are incorporated into the augmented pre-dual action.}
\begin{equation}
\begin{aligned}
	\mathcal{L}_{\mathrm a}
	&=-x\dot{\xi}_x-y\dot{\xi}_y-mv_x\dot{\eta}_x-mv_y\dot{\eta}_y
	-\xi_xv_x-\xi_yv_y+k\eta_xxr+k\eta_yxs\\
	&\quad +\lambda(y^2-r)+\mu(x^2-s)-H_{\mathrm a}(x,y,r,s,v_x,v_y)\,.
\end{aligned}
\end{equation}
Consider the quadratic auxiliary function $	H_{\mathrm a}=\frac{\alpha}{2}\big(x^2+y^2+r^2+s^2\big)+\frac{\beta}{2}\big(v_x^2+v_y^2\big)$, $\alpha,\beta>0$. Stationarity with respect to the enlarged set of primal variables gives us
\begin{subequations}\label{AugmentedDualToPrimal}
\begin{align}
	\alpha x&=-\dot{\xi}_x+k\eta_xr+k\eta_ys+2\mu x\,,\\
	\alpha y&=-\dot{\xi}_y+2\lambda y\,,\\
	\alpha r&=k\eta_xx-\lambda\,,\\
	\alpha s&=k\eta_yx-\mu\,,\\
	\beta v_x&=-m\dot{\eta}_x-\xi_x\,,\\
	\beta v_y&=-m\dot{\eta}_y-\xi_y\,.
\end{align}
\end{subequations}
The first four equations in \eqref{AugmentedDualToPrimal} form the linear system
\begin{equation}
	\begin{bmatrix}
		\alpha-2\mu & 0 & -k\eta_x & -k\eta_y\\
		0 & \alpha-2\lambda & 0 & 0\\
		-k\eta_x & 0 & \alpha & 0\\
		-k\eta_y & 0 & 0 & \alpha
	\end{bmatrix}
	\begin{bmatrix}
		x\\y\\r\\s
	\end{bmatrix}
	=-\begin{bmatrix}
		\dot{\xi}_x\\ \dot{\xi}_y\\ \lambda\\ \mu
	\end{bmatrix}\,.
\end{equation}
Its determinant is $\alpha(\alpha-2\lambda)\left[\alpha(\alpha-2\mu)-k^2\big(\eta_x^2+\eta_y^2\big)\right]$.
Therefore, whenever this determinant is nonzero, the augmented dual-to-primal map is explicit and reads
\begin{equation}
\begin{aligned}
	\widehat{x} &=-\frac{\alpha\,\dot{\xi}_x+k(\eta_x\lambda+\eta_y\mu)}
	{\alpha(\alpha-2\mu)-k^2(\eta_x^2+\eta_y^2)}\,,\qquad
	\widehat{y}=-\frac{\dot{\xi}_y}{\alpha-2\lambda}\,,\\
	\widehat{r} &=\frac{k\,\eta_x\,\widehat{x}-\lambda}{\alpha}\,,\qquad
	\widehat{s}=\frac{k\,\eta_y\,\widehat{x}-\mu}{\alpha}\,,\qquad
	\widehat{\mathbf{v}}=-\frac{1}{\beta}
	\big(m\dot{\boldsymbol{\eta}}+\boldsymbol{\xi}\big)\,.
\end{aligned}
\end{equation}
Substitution of this mapping into the augmented pre-dual Lagrangian gives a reduced dual Lagrangian depending only on $\boldsymbol{\xi}$, $\boldsymbol{\eta}$, $\lambda$, $\mu$,
$\dot{\boldsymbol{\xi}}$, and $\dot{\boldsymbol{\eta}}$.
Stationarity with respect to $\lambda$ and $\mu$ recovers the algebraic constraints $\widehat{y}^{\,2}-\widehat{r}=0$ and $\widehat{x}^{\,2}-\widehat{s}=0$, while stationarity with respect to $\boldsymbol{\xi}$ and $\boldsymbol{\eta}$ recovers the original equations of motion. Thus, the quadratic lifting converts the implicit nonlinear dual-to-primal problem into an explicit linear algebraic problem in the enlarged set of primal variables. By introducing additional auxiliary variables, this lifting procedure can be extended to polynomial nonlinearities of arbitrary degree. 
When the coefficient matrix is singular, the dual-to-primal equations may have multiple solutions or no solution. In the former case, an additional criterion is needed to select one of the solutions; this works for the dual variational problem, but recovering primal solutions can require additional care related to considerations of regularity.
\end{remark}

\subsection{A three-dimensional nonlinear curl force}

Let us apply the dual variational formulation to the following three-dimensional curl force that was considered in \citep{YavariGoriely2025Curl}
\begin{equation}
	\mathbf{F}(x,y,z)=-(yz,2xz,xy)\,.
\end{equation}
The Jacobian of the force field is
\begin{equation}
	D\mathbf{F}(x,y,z)=-
	\begin{bmatrix}
		0 & z & y\\
		2z & 0 & 2x\\
		y & x & 0
	\end{bmatrix}\,.
\end{equation}
We choose the base state $\bar{\mathbf{x}}=\bar{\mathbf{v}}=\mathbf{0}$ and write the dual variables as $\boldsymbol{\xi}=(\xi_x,\xi_y,\xi_z)$ and $\boldsymbol{\eta}=(\eta_x,\eta_y,\eta_z)$. For a general smooth function $H=H(\mathbf{x},\mathbf{v})$, the dual-to-primal equations \eqref{eq:DtPEquations} take the form
\begin{subequations}\label{eq:3DGeneralDualToPrimal}
\begin{align}
	\frac{\partial H}{\partial x} &=-\dot{\xi}_x+2z\eta_y+y\eta_z\,,\\
	\frac{\partial H}{\partial y} &=-\dot{\xi}_y+z\eta_x+x\eta_z\,,\\
	\frac{\partial H}{\partial z} &=-\dot{\xi}_z+y\eta_x+2x\eta_y\,,\\
	\frac{\partial H}{\partial v_x} &=-m\dot{\eta}_x-\xi_x\,,\\
	\frac{\partial H}{\partial v_y} &=-m\dot{\eta}_y-\xi_y\,,\\
	\frac{\partial H}{\partial v_z} &=-m\dot{\eta}_z-\xi_z\,.
\end{align}
\end{subequations}
Provided that these equations can be solved locally for $(x,y,z,v_x,v_y,v_z)$, they define the dual-to-primal mapping. To obtain an explicit representation, consider the quadratic function
\begin{equation}
	H(\mathbf{x},\mathbf{v})=\frac{\alpha}{2}\big(x^2+y^2+z^2\big)
	+\frac{\beta}{2}\big(v_x^2+v_y^2+v_z^2\big)\,,\qquad \alpha,\beta>0\,.
\end{equation}
The velocity part of the dual-to-primal mapping reads
\begin{equation}
	\widehat{\mathbf{v}}=-\frac{1}{\beta}\big(m\dot{\boldsymbol{\eta}}+\boldsymbol{\xi}\big)\,.
\end{equation}
The position equations can be written as
\begin{equation}\label{eq:3DPositionDualToPrimal}
	\begin{bmatrix}
		\alpha & -\eta_z & -2\eta_y\\
		-\eta_z & \alpha & -\eta_x\\
		-2\eta_y & -\eta_x & \alpha
	\end{bmatrix}
	\begin{bmatrix}
		x\\y\\z
	\end{bmatrix}
	=-
	\begin{bmatrix}
		\dot{\xi}_x\\ \dot{\xi}_y\\ \dot{\xi}_z
	\end{bmatrix}\,.
\end{equation}
Let
\begin{equation}
	\mathbf{A}(\boldsymbol{\eta})=
	\begin{bmatrix}
		\alpha & -\eta_z & -2\eta_y\\
		-\eta_z & \alpha & -\eta_x\\
		-2\eta_y & -\eta_x & \alpha
	\end{bmatrix}\,.
\end{equation}
Whenever $\det\mathbf{A}(\boldsymbol{\eta})\neq0$, the position part of the dual-to-primal mapping is given as $\widehat{\mathbf{x}}=-\mathbf{A}^{-1}(\boldsymbol{\eta})\,\dot{\boldsymbol{\xi}}$. Thus, the complete dual-to-primal mapping is written as
\begin{equation}
	\widehat{\mathbf{x}}=-\mathbf{A}^{-1}(\boldsymbol{\eta})\,\dot{\boldsymbol{\xi}}\,,
	\qquad
	\widehat{\mathbf{v}}=-\frac{1}{\beta}\big(m\dot{\boldsymbol{\eta}}+\boldsymbol{\xi}\big)\,.
\end{equation}
The corresponding reduced dual Lagrangian reads
\begin{equation}
	\mathcal{L}_{\mathrm d}=-\widehat{\mathbf{x}}\cdot\dot{\boldsymbol{\xi}}
	-m\widehat{\mathbf{v}}\cdot\dot{\boldsymbol{\eta}}-\boldsymbol{\xi}\cdot\widehat{\mathbf{v}}
	+\eta_x\,\widehat{y}\,\widehat{z}+2\eta_y\,\widehat{x}\,\widehat{z}+\eta_z\,\widehat{x}\,\widehat{y}
	-\frac{\alpha}{2}\big(\widehat{x}^2+\widehat{y}^2+\widehat{z}^2\big)
	-\frac{\beta}{2}\big(\widehat{v}_x^2+\widehat{v}_y^2+\widehat{v}_z^2\big)\,.
\end{equation}
Stationarity of the associated dual action recovers
\begin{equation}
\begin{dcases}
	\dot{\widehat{x}}=\widehat{v}_x\,, & \dot{\widehat{y}}=\widehat{v}_y\,,\quad
	\dot{\widehat{z}}=\widehat{v}_z\,,\\[1ex]
	m\,\dot{\widehat{v}}_x=-\widehat{y}\,\widehat{z}\,,&
	m\,\dot{\widehat{v}}_y=-2\,\widehat{x}\,\widehat{z}\,,\quad
	m\,\dot{\widehat{v}}_z=-\widehat{x}\,\widehat{y}\,,
\end{dcases}
\end{equation}
together with the prescribed initial conditions. In contrast to the two-dimensional example (without the quadratic lifting), the quadratic structure of the force makes the position part of the dual-to-primal mapping linear and hence explicitly solvable away from the locus $\det\mathbf{A}(\boldsymbol{\eta})=0$.

\subsection{The Ziegler column}

\begin{figure}[t!]
\centering
\includegraphics[width=0.35\textwidth]{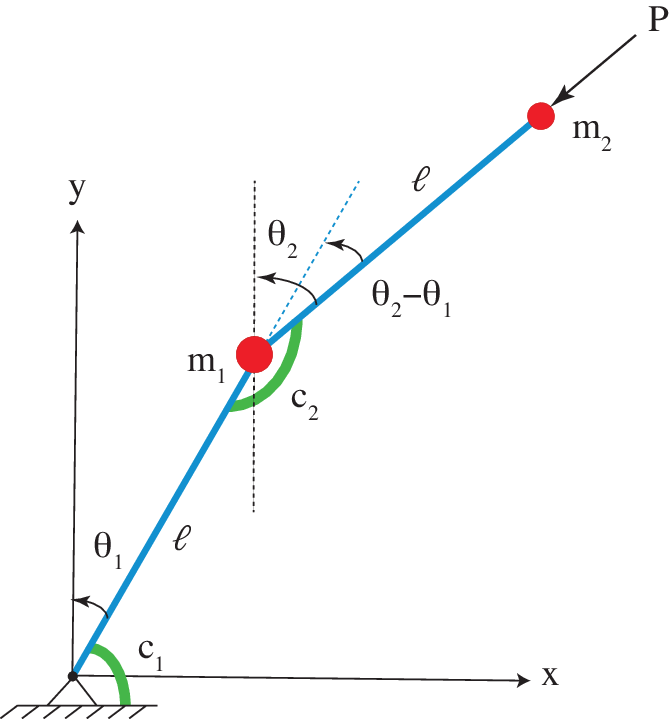}
\vspace*{0.10in}
\caption{The classical Ziegler column consisting of two massless rigid links of length $\ell$, point masses $m_1$ and $m_2$, rotational springs of stiffnesses $c_1$ and $c_2$, and a compressive follower force of magnitude $P$. The angles $\theta_1$ and $\theta_2$ are the absolute rotations of the two links measured from the vertical configuration, while $\theta_2-\theta_1$ is their relative rotation. The spring of stiffness $c_1$ is moment-free when $\theta_1=0$, while the spring of stiffness $c_2$ is moment-free when $\theta_2-\theta_1=0$; thus, both springs are moment-free in the initial vertical configuration.}
\label{Fig:Ziegler-Column}
\end{figure}

Let us consider the classical undamped Ziegler column, consisting of two massless rigid links of length $\ell$, carrying point masses $m_1$ and $m_2$, and connected by rotational springs of stiffnesses $c_1$ and $c_2$. A tangential follower force of magnitude $P$ is applied at the free end of the second link \citep{Ziegler1977,Kirillov2021}, see Fig.~\ref{Fig:Ziegler-Column}.

Let $\theta_1$ and $\theta_2$ denote the absolute rotations of the first and second links, respectively, measured from the straight equilibrium configuration. Thus, the relative rotation of the second spring is $\theta_2-\theta_1$. The position vectors of the two point masses are
\begin{equation}
	\mathbf{r}_1=\ell\big(\sin\theta_1,\cos\theta_1\big)\,,\qquad
	\mathbf{r}_2=\ell\big(\sin\theta_1+\sin\theta_2,\cos\theta_1+\cos\theta_2\big)\,.
\end{equation}
The two rigid links are assumed to be massless. Therefore, the kinetic energy of the system consists solely of the kinetic energies of the two point masses $m_1$ and $m_2$.
The velocity of the first mass is $\dot{\mathbf{r}}_1=\ell\big(\cos\theta_1\,\dot{\theta}_1,-\sin\theta_1\,\dot{\theta}_1\big)$, and hence $\dot{\mathbf{r}}_1\cdot\dot{\mathbf{r}}_1=\ell^2\dot{\theta}_1^2$. The velocity of the second mass is $\dot{\mathbf{r}}_2=\ell\big(\cos\theta_1\,\dot{\theta}_1+\cos\theta_2\,\dot{\theta}_2,
-\sin\theta_1\,\dot{\theta}_1-\sin\theta_2\,\dot{\theta}_2\big)$. Therefore,
\begin{equation}
\begin{aligned}
	\dot{\mathbf{r}}_2\cdot\dot{\mathbf{r}}_2
	=\ell^2\left[\dot{\theta}_1^2+\dot{\theta}_2^2
	+2\big(\cos\theta_1\cos\theta_2+\sin\theta_1\sin\theta_2\big)\,
	\dot{\theta}_1\dot{\theta}_2\right]
	=\ell^2\left[\dot{\theta}_1^2+\dot{\theta}_2^2
	+2\cos(\theta_1-\theta_2)\dot{\theta}_1\dot{\theta}_2\right]\,.
\end{aligned}
\end{equation}
Thus, the kinetic energy of the system is given as
\begin{equation}
	T=\frac{1}{2}m_1\dot{\mathbf{r}}_1\cdot\dot{\mathbf{r}}_1
	+\frac{1}{2}m_2\dot{\mathbf{r}}_2\cdot\dot{\mathbf{r}}_2
	=\frac{1}{2}m_1\ell^2\dot{\theta}_1^2
	+\frac{1}{2}m_2\ell^2\left[\dot{\theta}_1^2+\dot{\theta}_2^2
	+2\cos(\theta_1-\theta_2)\dot{\theta}_1\dot{\theta}_2\right]\,.
\end{equation}
Let $\mathbf{q}=(\theta_1,\theta_2)$ denote the vector of generalized coordinates. The kinetic energy can be written as\footnote{For $m_1,m_2>0$ and $\ell>0$, the symmetric mass matrix
$\mathbf{M}(\mathbf{q})$ is positive definite because its leading
principal minors are $\ell^2(m_1+m_2)>0$ and $\det\mathbf{M}(\mathbf{q})
=\ell^4m_2\big[m_1+m_2\sin^2(\theta_1-\theta_2)\big]>0$.}
\begin{equation}
	T=\frac{1}{2}\dot{\mathbf{q}}\cdot\mathbf{M}(\mathbf{q})\,\dot{\mathbf{q}}\,,\qquad
	\mathbf{M}(\mathbf{q})=\ell^2
	\begin{bmatrix}
		m_1+m_2 & m_2\cos(\theta_1-\theta_2)\\
		m_2\cos(\theta_1-\theta_2) & m_2
	\end{bmatrix}\,.
\end{equation}
The elastic energy stored in the two rotational springs is
\begin{equation}
	\Pi=\frac{1}{2}c_1\theta_1^2+\frac{1}{2}c_2(\theta_2-\theta_1)^2
	=\frac{1}{2}\mathbf{q}\cdot\mathbf{K}_{\mathrm e}\mathbf{q}\,,
	\qquad
	\mathbf{K}_{\mathrm e}=
	\begin{bmatrix}
		c_1+c_2 & -c_2\\
		-c_2 & c_2
	\end{bmatrix}\,.
\end{equation}
The follower force is applied at the second mass, has constant magnitude $P$, and remains tangent to the second link. Thus, $\mathbf{P}=-P\big(\sin\theta_2,\cos\theta_2\big)$. 
Its generalized forces are obtained from its virtual work:\footnote{Note that the virtual displacement of the second mass is written as $\delta\mathbf{r}_2=\frac{\partial\mathbf{r}_2}{\partial\theta_i}\delta\theta_i$, where summation over $i$ is understood. Therefore, $\delta W=\mathbf{P}\cdot\delta\mathbf{r}_2=\mathbf{P}\cdot\frac{\partial\mathbf{r}_2}{\partial\theta_i}\delta\theta_i$. Comparing this expression with the definition $\delta W=Q_i\delta\theta_i$ gives us $Q_i=\mathbf{P}\cdot\frac{\partial\mathbf{r}_2}{\partial\theta_i}$.}
\begin{equation}
	Q_i=\mathbf{P}\cdot\frac{\partial\mathbf{r}_2}{\partial\theta_i}\,,\qquad i=1,2\,.
\end{equation}
A direct calculation gives us $Q_1=P\ell\sin(\theta_1-\theta_2)$, and $Q_2=0$. The second generalized force vanishes because the displacement of the end point produced by varying $\theta_2$ is perpendicular to the second link and, therefore, perpendicular to the follower force. In contrast, varying $\theta_1$ moves the entire second link and produces nonzero work.

To obtain the classical linearized Ziegler column, we linearize the equations of motion about the vertical configuration $\theta_1=\theta_2=0$. The mass matrix evaluated at this configuration is
\begin{equation}
	\mathbf{M}=\mathbf{M}(\mathbf{0})=\ell^2
	\begin{bmatrix}
		m_1+m_2 & m_2\\
		m_2 & m_2
	\end{bmatrix}\,,
\end{equation}
and the quadratic kinetic energy of the linearized system is $T=\frac{1}{2}\dot{\mathbf{q}}\cdot\mathbf{M}\dot{\mathbf{q}}$. 
Linearization of the generalized follower force gives us
\begin{equation}
	\mathbf{Q}_{\mathrm f}=P\ell
	\begin{bmatrix}
		1 & -1\\
		0 & 0
	\end{bmatrix}\mathbf{q}
	=\mathbf{K}_{\mathrm f}\mathbf{q}\,.
\end{equation}
The linearized equations of motion are consequently written as $\mathbf{M}\ddot{\mathbf{q}}+\mathbf{K}_{\mathrm e}\mathbf{q}=\mathbf{K}_{\mathrm f}\mathbf{q}$, or, equivalently,
\begin{equation}\label{eq:ZieglerEquations}
	\mathbf{M}\ddot{\mathbf{q}}+\mathbf{K}\mathbf{q}=\mathbf{0}\,,
	\qquad
	\mathbf{K}=\mathbf{K}_{\mathrm e}-\mathbf{K}_{\mathrm f}
	=\begin{bmatrix}
		c_1+c_2-P\ell & P\ell-c_2\\
		-c_2 & c_2
	\end{bmatrix}\,.
\end{equation}
Notice that $\mathbf{K}_{\mathrm e}$ is symmetric, whereas $\mathbf{K}_{\mathrm f}$ and hence $\mathbf{K}$ are nonsymmetric. This asymmetry is a direct consequence of the nonconservative follower force: the generalized force $Q_1$ depends on both $\theta_1$ and $\theta_2$, while $Q_2$ vanishes.
The generalized force in \eqref{eq:ZieglerEquations} is $\mathbf{F}(\mathbf{q})=-\mathbf{K}\mathbf{q}$. Since
\begin{equation}
	\frac{\partial F_2}{\partial\theta_1}-\frac{\partial F_1}{\partial\theta_2}
	=K_{12}-K_{21}=P\ell\,,
\end{equation}
the follower load generates a curl force whenever $P\neq0$.

Let $\mathbf{v}=\dot{\mathbf{q}}$, and denote the dual variables by $\boldsymbol{\xi}=(\xi_1,\xi_2)$ and $\boldsymbol{\eta}=(\eta_1,\eta_2)$. For a general smooth function $H=H(\mathbf{q},\mathbf{v})$, the pre-dual Lagrangian reads
\begin{equation}
	\mathcal{L}=-\mathbf{q}\cdot\dot{\boldsymbol{\xi}}
	-\mathbf{M}\mathbf{v}\cdot\dot{\boldsymbol{\eta}}
	-\boldsymbol{\xi}\cdot\mathbf{v}
	+\boldsymbol{\eta}\cdot\mathbf{K}\mathbf{q}
	-H(\mathbf{q},\mathbf{v})\,.
\end{equation}
Stationarity with respect to $\mathbf{q}$ and $\mathbf{v}$ gives the dual-to-primal equations
\begin{equation}\label{eq:ZieglerGeneralDualToPrimal}
	\frac{\partial H}{\partial\mathbf{q}}=-\dot{\boldsymbol{\xi}}+\mathbf{K}^{\mathsf T}\boldsymbol{\eta}\,,
	\qquad
	\frac{\partial H}{\partial\mathbf{v}}=-\mathbf{M}^{\mathsf T}\dot{\boldsymbol{\eta}}-\boldsymbol{\xi}
	=-\mathbf{M}\dot{\boldsymbol{\eta}}-\boldsymbol{\xi}\,,
\end{equation}
where the symmetry of the mass matrix $\mathbf{M}$ has been used.
Provided that \eqref{eq:ZieglerGeneralDualToPrimal} can be solved locally for $(\mathbf{q},\mathbf{v})$, it defines the dual-to-primal mapping.

In order to obtain an explicit representation, consider the following quadratic function
\begin{equation}
	H(\mathbf{q},\mathbf{v})=\frac{1}{2}\mathbf{q}\cdot\mathbf{A}\mathbf{q}
	+\frac{1}{2}\mathbf{v}\cdot\mathbf{B}\mathbf{v}\,,
\end{equation}
where $\mathbf{A}$ and $\mathbf{B}$ are symmetric positive-definite matrices. The dual-to-primal mapping is then
\begin{equation}\label{eq:ZieglerDualToPrimal}
	\widehat{\mathbf{q}}=\mathbf{A}^{-1}
	\big(\mathbf{K}^{\mathsf T}\boldsymbol{\eta}-\dot{\boldsymbol{\xi}}\big)\,,
	\qquad
	\widehat{\mathbf{v}}=-\mathbf{B}^{-1}
	\big(\mathbf{M}\dot{\boldsymbol{\eta}}+\boldsymbol{\xi}\big)\,.
\end{equation}
Substitution of \eqref{eq:ZieglerDualToPrimal} into the pre-dual Lagrangian gives the reduced dual Lagrangian
\begin{equation}
	\mathcal{L}_{\mathrm d}
	=\frac{1}{2}\big(\mathbf{K}^{\mathsf T}\boldsymbol{\eta}-\dot{\boldsymbol{\xi}}\big)
	\cdot\mathbf{A}^{-1}\big(\mathbf{K}^{\mathsf T}\boldsymbol{\eta}-\dot{\boldsymbol{\xi}}\big)
	+\frac{1}{2}\big(\mathbf{M}\dot{\boldsymbol{\eta}}+\boldsymbol{\xi}\big)
	\cdot\mathbf{B}^{-1}\big(\mathbf{M}\dot{\boldsymbol{\eta}}+\boldsymbol{\xi}\big)\,.
\end{equation}
Thus, the dual action is written as
\begin{equation}\label{eq:ZieglerDualAction}
	S[\boldsymbol{\xi},\boldsymbol{\eta}]
	=\int_0^T\mathcal{L}_{\mathrm d}\,\mathrm{d}t
	-\mathbf{q}_0\cdot\boldsymbol{\xi}(0)
	-\mathbf{M}\mathbf{v}_0\cdot\boldsymbol{\eta}(0)\,,
\end{equation}
where the terminal values $\boldsymbol{\xi}(T)$ and $\boldsymbol{\eta}(T)$ are prescribed. Stationarity of \eqref{eq:ZieglerDualAction} recovers
\begin{equation}\label{eq:dual_lin_ziegler}
	\dot{\widehat{\mathbf{q}}}=\widehat{\mathbf{v}}\,,\qquad
	\mathbf{M}\dot{\widehat{\mathbf{v}}}+\mathbf{K}\widehat{\mathbf{q}}=\mathbf{0}\,,
	\qquad
	\widehat{\mathbf{q}}(0)=\mathbf{q}_0\,,\qquad
	\widehat{\mathbf{v}}(0)=\mathbf{v}_0\,.
\end{equation}

\begin{remark}
A convenient choice is $\mathbf{A}=\mathbf{I}$ and
$\mathbf{B}=\mathbf{M}^{2}$. Notice that the coefficients of $\ddot{\boldsymbol{\xi}}$ and $\ddot{\boldsymbol{\eta}}$ in the dual
Euler--Lagrange equations are $\mathbf{A}^{-1}$ and
$\mathbf{M}\mathbf{B}^{-1}\mathbf{M}$, respectively. With the above choices, both coefficients reduce to the identity. Thus, the principal part of the dual Euler--Lagrange system has the identity as its coefficient matrix, making its elliptic character explicit and eliminating coupling between highest-order derivatives.
\end{remark}

Thus, although the nonsymmetric stiffness matrix of the Ziegler column prevents the existence of an ordinary elastic potential, its initial-value problem
admits an explicit quadratic dual action. In particular, the choices $\mathbf{A}=\mathbf{I}$ and $\mathbf{B}=\mathbf{M}^{2}$ reduce the coefficients of $\ddot{\boldsymbol{\xi}}$ and $\ddot{\boldsymbol{\eta}}$ in the dual Euler--Lagrange equations to the identity. Thus, the principal part of the dual Euler--Lagrange system is an identity-coefficient, uniformly elliptic second-order operator. The circulatory character of the follower force enters the dual formulation through the transpose $\mathbf{K}^{\mathsf T}$ in the dual-to-primal mapping. Nevertheless, the second variation of the quadratic dual functional is symmetric. Consequently, discretizations constructed directly from the dual functional inherit this symmetry and lead to
symmetric discrete algebraic systems. The resulting dual problem can be solved directly on the full time interval. For long time intervals, the
time-staging procedures introduced in \citep{ka2} provide a more efficient alternative by replacing a single large global system with a sequence of smaller problems.

The corresponding auxiliary dual Hamiltonian reads
\begin{equation}
	\mathscr{H}_{\mathrm d}
	=\boldsymbol{\xi}\cdot\widehat{\mathbf{v}}
	-\boldsymbol{\eta}\cdot\mathbf{K}\widehat{\mathbf{q}}
	+\frac{1}{2}\widehat{\mathbf{q}}\cdot\mathbf{A}\widehat{\mathbf{q}}
	+\frac{1}{2}\widehat{\mathbf{v}}\cdot\mathbf{B}\widehat{\mathbf{v}}\,,
\end{equation}
and is conserved along extremal dual trajectories, and therefore also along solutions of the primal system \eqref{eq:dual_lin_ziegler}. The dual dynamics of this linear curl-force system therefore admits a global single-valued Hamiltonian.

\begin{remark}
More generally, any dual Lagrangian for the linear primal dynamics for $t \mapsto \mathbf{x}(t) \in \R^n$, $n$ a positive integer, of the type $\mathbf{M} \ddot{\mathbf{x}} + \mathbf{D} \dot{\mathbf{x}} + \mathbf{K} \mathbf{x}  = \mathbf{0}$, $\mathbf{x}(0) = \mathbf{x}_0$, $\dot{\mathbf{x}}(0) = \mathbf{v}_0$, with given $\mathbf{x}_0, \mathbf{v}_0 \in \R^n$, $\mathbf{M} \in \R^{n \times n}_{sym}$ and invertible,  and $\mathbf{K}, \mathbf{D} \in \R^{n \times n}$ without any further special properties, admits a global single-valued dual Hamiltonian when the dual Lagrangian is parametrized by two $\R^{n \times n}$ symmetric positive definite matrices $\mathbf{A}, \mathbf{B}$ appearing in the auxiliary potential $H(\mathbf{x}, \mathbf{v}) = \half \left( \mathbf{x} \cdot \mathbf{A} \mathbf{x} + \mathbf{v} \cdot \mathbf{B} \mathbf{v}\right)$.  The primal dynamics includes conservative, non-conservative (e.g. curl), as well as dissipative forces.
\end{remark}

\section{Conclusions}  \label{Sec:Conclusions}

In this paper, we presented a dual variational formulation for the dynamics of particles under curl forces. Although such forces do not possess an ordinary potential energy and their equations of motion do not, in general, follow from a standard variational principle, we showed that their initial-value problems can be obtained from an action expressed entirely in terms of dual variables. The construction begins with a pre-dual Lagrangian involving the primal and dual states and an auxiliary function $H$. Stationarity with respect to the primal variables defines a local dual-to-primal mapping, whose substitution into the pre-dual action gives the dual action. The Euler--Lagrange equations of this action recover both the original equations of motion and the prescribed initial conditions.

The freedom in the choice of $H$ generates a family of dual variational formulations for the same curl-force dynamics. This function does not represent a physical energy; rather, it controls the structure and local invertibility of the dual-to-primal mapping. We also constructed an auxiliary dual Hamiltonian that is conserved along stationary dual trajectories. This conserved quantity is associated with the dual variational structure and, in general, is distinct from the physical energy of the particle.

The formulation was illustrated using nonlinear curl forces in two and three dimensions and the classical Ziegler column. The nonlinear examples showed that the nonlinearity of the force is transferred to a pointwise algebraic dual-to-primal problem. For the linearized Ziegler column, the mapping is explicit and the dual action is quadratic, despite the nonsymmetry of the circulatory stiffness matrix. These examples
demonstrate that the absence of an ordinary potential energy does not preclude a variational description of curl-force dynamics. Building on the works of Acharya and co-workers already mentioned, extensions of the present formulation to systems of particles, continuum theories with non-conservative constitutive responses, and numerical schemes based on dual actions remain topics for future work. Another promising direction is the computation of periodic orbits in possibly chaotic curl-force systems, such as those considered in \citep{Berry2025Illustrations}, for which a variational formulation is particularly well suited.



\bibliographystyle{abbrvnat}
\bibliography{ref}

\begin{thebibliography}{44}
\providecommand{\natexlab}[1]{#1}
\providecommand{\url}[1]{\texttt{#1}}
\expandafter\ifx\csname urlstyle\endcsname\relax
  \providecommand{\doi}[1]{doi: #1}\else
  \providecommand{\doi}{doi: \begingroup \urlstyle{rm}\Url}\fi

\bibitem[Acharya(2023{\natexlab{a}})]{ach2}
A.~Acharya.
\newblock A dual variational principle for nonlinear dislocation dynamics.
\newblock \emph{Journal of Elasticity}, 154\penalty0 (1):\penalty0 383--395,
  2023{\natexlab{a}}.

\bibitem[Acharya(2023{\natexlab{b}})]{acharyaQAM}
A.~Acharya.
\newblock Variational principles for nonlinear {PDE} systems via duality.
\newblock \emph{Quarterly of Applied Mathematics}, LXXXI\penalty0 (1):\penalty0
  127--140, 2023{\natexlab{b}}.

\bibitem[Acharya(2025{\natexlab{a}})]{Achmhd}
A.~Acharya.
\newblock Ideal magnetohydrodynamics and field dislocation mechanics.
\newblock \emph{Pure and Applied Functional Analysis}, 10\penalty0
  (1):\penalty0 1--10, 2025{\natexlab{a}}.
\newblock ISSN 2189-3756,2189-3764.

\bibitem[Acharya(2025{\natexlab{b}})]{ach3}
A.~Acharya.
\newblock A hidden convexity in continuum mechanics, with application to
  classical, continuous-time, rate-(in)dependent plasticity.
\newblock \emph{Mathematics and Mechanics of Solids}, 30(3):\penalty0 701--719,
  2025{\natexlab{b}}.
\newblock URL \url{https://arxiv.org/abs/2310.03201}.

\bibitem[Acharya(2026)]{acharya2026new}
A.~Acharya.
\newblock A new perspective in linear {Cauchy Elasticity}: variational minimum
  principles for statics, dynamics, and heterogeneous materials.
\newblock \emph{arXiv preprint}, 2026.
\newblock URL \url{https://arxiv.org/abs/2606.24782}.

\bibitem[{Acharya} and {Ginster}(2025)]{AG_control}
A.~{Acharya} and J.~{Ginster}.
\newblock {A convex variational principle for the necessary conditions of
  classical optimal control}.
\newblock \emph{arXiv e-prints}, 2025.
\newblock URL \url{https://arxiv.org/abs/2502.15973}.

\bibitem[Acharya and Sengupta(2024{\natexlab{a}})]{Ach6}
A.~Acharya and A.~N. Sengupta.
\newblock Action principles for dissipative, non-holonomic {N}ewtonian
  mechanics.
\newblock \emph{Proceedings of the Royal Society A.}, 480\penalty0
  (2293):\penalty0 Paper No. 20240113, 21, 2024{\natexlab{a}}.

\bibitem[Acharya and Sengupta(2024{\natexlab{b}})]{ach_cmds}
A.~Acharya and A.~N. Sengupta.
\newblock Variational principle for a damped, quadratically interacting
  particle chain with nonconservative forcing.
\newblock In F.~Willot, J.~Dirrenberger, S.~Forest, D.~Jeulin, and A.~V.
  Cherkaev, editors, \emph{Continuum Models and Discrete Systems}, pages
  195--201, Cham, 2024{\natexlab{b}}. Springer Nature Switzerland.
\newblock ISBN 978-3-031-58665-1.

\bibitem[{Acharya} et~al.(2024{\natexlab{a}}){Acharya}, {Ginster}, and
  {Sengupta}]{AGS24}
A.~{Acharya}, J.~{Ginster}, and A.~N. {Sengupta}.
\newblock {Variational Dual Solutions of Chern-Simons Theory}.
\newblock \emph{arXiv e-prints}, Nov. 2024{\natexlab{a}}.
\newblock URL \url{https://arxiv.org/abs/2411.17635}.

\bibitem[{Acharya} et~al.(2024{\natexlab{b}}){Acharya}, {Stroffolini}, and
  {Zarnescu}]{ASZ24}
A.~{Acharya}, B.~{Stroffolini}, and A.~{Zarnescu}.
\newblock {Variational Dual Solutions for Incompressible Fluids}.
\newblock \emph{arXiv e-prints}, Sept. 2024{\natexlab{b}}.
\newblock URL \url{https://arxiv.org/abs/2409.04911}.

\bibitem[Beck(1952)]{Beck1952}
M.~Beck.
\newblock Die {K}nicklast des einseitig eingespannten, tangential
  gedr{\"u}ckten {S}tabes.
\newblock \emph{Zeitschrift f{\"u}r angewandte Mathematik und Physik},
  3:\penalty0 225--228, 1952.

\bibitem[Berry(2025)]{Berry2025Illustrations}
M.~V. Berry.
\newblock Six illustrations of curl force dynamics.
\newblock \emph{European Journal of Physics}, 46\penalty0 (6):\penalty0 065003,
  2025.

\bibitem[Berry and Shukla(2012)]{Berry2012}
M.~V. Berry and P.~Shukla.
\newblock Classical dynamics with curl forces, and motion driven by
  time-dependent flux.
\newblock \emph{Journal of Physics A: Mathematical and Theoretical},
  45\penalty0 (30):\penalty0 305201, 2012.

\bibitem[Berry and Shukla(2013)]{Berry2013}
M.~V. Berry and P.~Shukla.
\newblock Physical curl forces: dipole dynamics near optical vortices.
\newblock \emph{Journal of Physics A: Mathematical and Theoretical},
  46\penalty0 (42):\penalty0 422001, 2013.

\bibitem[Berry and Shukla(2015)]{Berry2015}
M.~V. Berry and P.~Shukla.
\newblock Hamiltonian curl forces.
\newblock \emph{Proceedings of the Royal Society A}, 471\penalty0
  (2176):\penalty0 20150002, 2015.

\bibitem[Bigoni and Noselli(2011)]{Bigoni2011}
D.~Bigoni and G.~Noselli.
\newblock Experimental evidence of flutter and divergence instabilities induced
  by dry friction.
\newblock \emph{Journal of the Mechanics and Physics of Solids}, 59\penalty0
  (10):\penalty0 2208--2226, 2011.

\bibitem[Bigoni et~al.(2018)Bigoni, Kirillov, Misseroni, Noselli, and
  Tommasini]{Bigoni2018}
D.~Bigoni, O.~N. Kirillov, D.~Misseroni, G.~Noselli, and M.~Tommasini.
\newblock Flutter and divergence instability in the {P}fl{\"u}ger column:
  {E}xperimental evidence of the {Z}iegler destabilization paradox.
\newblock \emph{Journal of the Mechanics and Physics of Solids}, 116:\penalty0
  99--116, 2018.

\bibitem[Bolotin(1963)]{Bolotin1963}
V.~V. Bolotin.
\newblock \emph{Nonconservative Problems of the Theory of Elastic Stability}.
\newblock Pergamon Press, London, 1963.

\bibitem[Cazzolli et~al.(2020)Cazzolli, Dal~Corso, and Bigoni]{Cazzolli2020}
A.~Cazzolli, F.~Dal~Corso, and D.~Bigoni.
\newblock Non-holonomic constraints inducing flutter instability in structures
  under conservative loadings.
\newblock \emph{Journal of the Mechanics and Physics of Solids}, 138:\penalty0
  103919, 2020.

\bibitem[Elishakoff(2005)]{Elishakoff2005}
I.~Elishakoff.
\newblock Controversy associated with the so-called ``follower forces":
  {C}ritical overview.
\newblock \emph{Applied Mechanics Reviews}, 58\penalty0 (2):\penalty0 117--142,
  2005.

\bibitem[Galley(2013)]{gall}
C.~R. Galley.
\newblock Classical mechanics of nonconservative systems.
\newblock \emph{Physical Review Letters}, 110\penalty0 (17):\penalty0 174301,
  2013.

\bibitem[Ghose-Choudhury and Guha(2019)]{GhoseChoudhuryGuha2019}
A.~Ghose-Choudhury and P.~Guha.
\newblock Hamiltonian description of nonlinear curl forces from cofactor
  systems.
\newblock \emph{Acta Mechanica}, 230\penalty0 (6):\penalty0 2267--2277, 2019.
\newblock \doi{10.1007/s00707-019-02394-y}.

\bibitem[Guha(2018)]{Guha2018}
P.~Guha.
\newblock Saddle in linear curl forces, cofactor systems and holomorphic
  structure.
\newblock \emph{The European Physical Journal Plus}, 133\penalty0
  (12):\penalty0 536, 2018.
\newblock \doi{10.1140/epjp/i2018-12341-2}.

\bibitem[Guha(2020)]{Guha2020}
P.~Guha.
\newblock Curl forces and their role in optics and ion trapping.
\newblock \emph{The European Physical Journal D}, 74:\penalty0 99, 2020.
\newblock \doi{10.1140/epjd/e2020-100462-6}.

\bibitem[Gurtin(1964)]{gurt}
M.~E. Gurtin.
\newblock Variational principles for linear initial-value problems.
\newblock \emph{Quarterly of Applied Mathematics}, XXII\penalty0 (3):\penalty0
  252--256, 1964.

\bibitem[Kirillov(2021)]{Kirillov2021}
O.~N. Kirillov.
\newblock \emph{Nonconservative Stability Problems of Modern Physics},
  volume~14.
\newblock Walter de Gruyter GmbH \& Co KG, 2021.

\bibitem[Koiter(1996)]{Koiter1996}
W.~T. Koiter.
\newblock Unrealistic follower forces.
\newblock \emph{Journal of Sound and Vibration}, 194\penalty0 (4):\penalty0
  636, 1996.

\bibitem[Kouskiya and Acharya(2024)]{ka1}
U.~Kouskiya and A.~Acharya.
\newblock Hidden convexity in the heat, linear transport, and {E}uler's rigid
  body equations: {A} computational approach.
\newblock \emph{Quarterly of Applied Mathematics}, LXXXII:\penalty0 673--703,
  2024.

\bibitem[Kouskiya and Acharya(2025)]{ka2}
U.~Kouskiya and A.~Acharya.
\newblock {I}nviscid {B}urgers as a degenerate elliptic problem.
\newblock \emph{Quarterly of Applied Mathematics}, LXXXIII:\penalty0 315--360,
  2025.

\bibitem[Kouskiya et~al.(2025)Kouskiya, Pego, and Acharya]{kpa}
U.~Kouskiya, R.~L. Pego, and A.~Acharya.
\newblock Traveling wave profiles for a semi-discrete {B}urgers equation.
\newblock \emph{Physica D}, page 134961, 2025.
\newblock URL \url{https://arxiv.org/abs/2504.12171}.

\bibitem[Kycia(2025)]{Kycia2025Classification}
R.~A. Kycia.
\newblock Classification of curl forces for all space dimensions.
\newblock \emph{arXiv preprint arXiv:2507.09817}, 2025.

\bibitem[Lundmark(2003)]{Lundmark2003}
H.~Lundmark.
\newblock Higher-dimensional integrable {N}ewton systems with quadratic
  integrals of motion.
\newblock \emph{Studies in Applied Mathematics}, 110\penalty0 (3):\penalty0
  257--296, 2003.
\newblock \doi{10.1111/1467-9590.00239}.

\bibitem[Pfl{\"u}ger(1950)]{Pfluger1950}
A.~Pfl{\"u}ger.
\newblock \emph{Stabilit{\"a}tsprobleme der Elastostatik}.
\newblock Springer, Berlin-G\"ottingen-Heidelberg, 1950.

\bibitem[Pfl{\"u}ger(1955)]{Pfluger1955}
A.~Pfl{\"u}ger.
\newblock Zur {S}tabilit{\"a}t des tangential gedr{\"u}ckten {S}tabes.
\newblock \emph{Zeitschrift f{\"u}r Angewandte Mathematik und Mechanik},
  35\penalty0 (5):\penalty0 191--191, 1955.

\bibitem[Rauch-Wojciechowski et~al.(1999)Rauch-Wojciechowski, Marciniak, and
  Lundmark]{RauchWojciechowskiEtAl1999}
S.~Rauch-Wojciechowski, K.~Marciniak, and H.~Lundmark.
\newblock Quasi-{L}agrangian systems of {N}ewton equations.
\newblock \emph{Journal of Mathematical Physics}, 40\penalty0 (12):\penalty0
  6366--6398, 1999.
\newblock \doi{10.1063/1.533098}.

\bibitem[Rothkopf and Horowitz(2026)]{rothkopf2026variational}
A.~Rothkopf and W.~Horowitz.
\newblock Variational approach to nonholonomic and inequality-constrained
  mechanics.
\newblock \emph{Physical Review E}, 113\penalty0 (2):\penalty0 024126, 2026.

\bibitem[Singh et~al.(2024)Singh, Ginster, and Acharya]{sga}
S.~Singh, J.~Ginster, and A.~Acharya.
\newblock A hidden convexity of nonlinear elasticity.
\newblock \emph{Journal of Elasticity}, 156\penalty0 (3):\penalty0 975--1014,
  2024.

\bibitem[Sukumar and Acharya(2025)]{suku_ach}
N.~Sukumar and A.~Acharya.
\newblock Variational formulation based on duality to solve partial
  differential equations: Use of {B}-splines and machine learning approximants.
\newblock \emph{Computer Methods in Applied Mechanics and Engineering},
  441:\penalty0 117909, 2025.

\bibitem[Vorotnikov and Acharya(2025)]{AV_nash}
D.~Vorotnikov and A.~Acharya.
\newblock On the variational dual formulation of the {N}ash system and an
  adaptive convex gradient-flow approach to nonlinear {PDE}s.
\newblock \emph{arXiv e-prints}, 2025.
\newblock URL \url{https://arxiv.org/abs/2512.12878}.

\bibitem[Yavari and Goriely(2025{\natexlab{a}})]{YavariGoriely2025Cauchy}
A.~Yavari and A.~Goriely.
\newblock Nonlinear {C}auchy elasticity.
\newblock \emph{Archive for Rational Mechanics and Analysis}, 249\penalty0
  (5):\penalty0 57, 2025{\natexlab{a}}.

\bibitem[Yavari and Goriely(2025{\natexlab{b}})]{YavariGoriely2025Curl}
A.~Yavari and A.~Goriely.
\newblock The {D}arboux classification of curl forces.
\newblock \emph{Journal of Physics A: Mathematical and Theoretical},
  58\penalty0 (27):\penalty0 275701, 2025{\natexlab{b}}.

\bibitem[Ziegler(1952)]{Ziegler1952}
H.~Ziegler.
\newblock Die {S}tabilit{\"a}tskriterien der {E}lastomechanik.
\newblock \emph{Ingenieur-Archiv}, 20\penalty0 (1):\penalty0 49--56, 1952.

\bibitem[Ziegler(1953)]{Ziegler1953}
H.~Ziegler.
\newblock Linear elastic stability: {A} critical analysis of methods.
\newblock \emph{Zeitschrift f{\"u}r angewandte Mathematik und Physik},
  4\penalty0 (3):\penalty0 167--185, 1953.

\bibitem[Ziegler(1977)]{Ziegler1977}
H.~Ziegler.
\newblock \emph{Principles of Structural Stability}.
\newblock Springer, Basel, 1977.

\end{thebibliography}

\end{document}